\documentstyle[12pt]{article}

\begin{document}
\begin{titlepage}
\begin{center}
{\Large\bf A new study of the polarized parton\\
\vskip 0.5cm
densities in the nucleon }
\end{center}
\vskip 2cm
\begin{center}
{\bf Elliot Leader}\\
{\it Birkbeck College, University of London\\
Malet Street, London WC1E 7HX, England\\
E-mail: e.leader@physics.bbk.ac.uk}\\
\vskip 0.5cm
{\bf Aleksander V. Sidorov}\\
{\it Bogoliubov Theoretical Laboratory\\
Joint Institute for Nuclear Research\\
41980 Dubna, Russia\\
E-mail: sidorov@thsun1.jinr.ru}
\vskip 0.5cm
{\bf Dimiter B. Stamenov \\
{\it Institute for Nuclear Research and Nuclear Energy\\
Bulgarian Academy of Sciences\\
blvd. Tsarigradsko Chaussee 72, Sofia 1784, Bulgaria\\
E-mail:stamenov@inrne.bas.bg }}
\end{center}

\vskip 0.3cm
\begin{abstract}

We present a new next-to-leading order QCD analysis of the world
data on inclusive polarized deep inelastic lepton-nucleon
scattering adding to the old set of data the final SMC results,
the HERMES proton and very recent SLAC/E155 deuteron data. We
find an excellent fit to the data and present results for the
polarized parton densities in different factorization schemes.
These results are in a good agreement with what follows from 
the theory.
We have also found that the main effect of the newly incorporated 
data is a better determination of the polarized gluon density.\\

\end{abstract}
\vskip 0.5 cm
\end{titlepage}
\newpage
\setcounter{page}{1}

\vskip 4mm
Recently \cite{LSiSt98, schemedep} we carried out a
next-to-leading order (NLO) QCD
analysis of the world data on polarized inclusive deep inelastic 
scattering (DIS). Since then the SMC group at CERN have re-analyzed 
their small-x data, using a totally new method of analysis, and 
published the
final results of their experiment \cite{finalSMC}. At the same
time new data have emerged from the HERMES Collaboration at DESY
\cite{HERMESp} and very recently from the E155 experiment at SLAC
\cite{E155d}.\\

The SMC group have themselves carried out an NLO QCD analysis 
\cite{SMC/QCD} based on their final results and most of the 
world data available at that time (with the
exception of the final results from the SLAC/E143 experiment
\cite{finSLACpd}). Given that there are significant changes in the
SMC data, and the high precision of the new HERMES and 
SLAC/E155 data, we felt it necessary to redo our analysis.\\

We have found that the newly incorporated data improve
the determination of the polarized parton densities, in
particular, of the poorly known gluon density. In this letter
we present a brief summary of the general trends in our new results
as well as a concise description of our present knowledge of the
polarized parton densities. \\

All details of our approach are given in \cite{LSiSt98, schemedep}.
Here we simply recall that we carry out our analysis in three 
different factorization schemes:$\overline{\rm MS}$, 
AB (Adler-Bardeen) \cite{ABscheme} and JET \cite{JETscheme}. 
We then test the stability of our analysis by comparing densities 
in one scheme, as determined from the fit to the data, with these 
same densities evaluated using the transformation rules connecting
densities in different schemes.  

We prefer to work with separate valence and sea parton densities
and we check the stability of our results when we vary the parameter
$\lambda$ corresponding to different flavour decompositions of the sea
\begin{equation}
\Delta\bar{u}=\Delta\bar{d}=\lambda \Delta\bar{s}~,
\label{SU3br}
\end{equation}
i.e., we confirm that the non-singlets $\Delta q_3,~\Delta q_8,~$
the singlet $\Delta \Sigma$ and the gluon density
$\Delta G$ are invariant. Note that this implies that the strange
quark density
\begin{equation}
\Delta\bar{s}={1\over 6}(\Delta \Sigma - \Delta q_8)~,
\label{s}
\end{equation}
is also invariant (does {\it not} change as $\lambda$ is varied) 
and it can be extracted from the data as well as $\Delta \Sigma$.
Of course, the valence quark densities $\Delta u_v$ and
$\Delta d_v$ are sensitive to the different assumptions about the
sea. Nonetheless they are of interest for predicting the
behaviour of other processes, e.g., polarized semi-inclusive DIS,
polarized Drell-Yan reactions, etc.\\

For the input polarized parton densities at $Q^2_0=1~GeV^2$ we have 
adopted a very simple parametrization 
\begin{eqnarray}
\nonumber
\Delta u_v(x,Q^2_0)&=&\eta_u A_ux^{a_u}xu_v(x,Q^2_0)~,\\
\nonumber
\Delta d_v(x,Q^2_0)&=&\eta_d A_dx^{a_d}xd_v(x,Q^2_0)~,\\
\nonumber
\Delta Sea(x,Q^2_0)&=&\eta_S A_Sx^{a_S}xSea(x,Q^2_0)~,\\
\Delta G(x,Q^2_0)&=&\eta_g A_gx^{a_g}xG(x,Q^2_0)~,
\label{classic}
\end{eqnarray}
where on RHS of (\ref{classic}) we have used the MRST 
unpolarized densities \cite{MRST}. The normalization factors 
$~A_f~$ in (\ref{classic}) are fixed such that $~\eta_{f}$ are the first 
moments of the polarized densities.

The first moments of the valence quark  densities
$~\eta_u~$ and $~\eta_d~$ are fixed by the octet nucleon and 
hyperon $\beta$ decay constants \cite{PDG}
\begin{equation}
g_{A}=F+D=1.2573~\pm~0.0028,~~~a_8=3F-D=0.579~\pm~0.025~,
\label{GA3FD}
\end{equation}
and in the case of SU(3) flavour symmetry of the sea  
($\Delta\bar{u}=\Delta\bar{d}=\Delta\bar{s}~$ at $~Q^2_0$)
\begin{equation}
\eta_u=0.918~,~~~~~~~~\eta_d=-0.339~.
\label{etaudSU3}
\end{equation}

The rest of the parameters in (\ref{classic}),
\begin{equation}
\{a_u,~a_d,~\eta_S,~a_S~,~\eta_g~,a_g\}~,
\label{claspar}
\end{equation}
have been determined from the best fit to the $~A_1^N(x,Q^2)~$ data.\\

The numerical results of our fits to the world data on 
$~A_1^N(x,Q^2)~$ [3-5, 7, 12-15] 
are summarized in Table 1.  
The data used (161 experimental points) cover the following 
kinematic region:  
\begin{equation}
0.004< x < 0.75,~~~~~~1< Q^2< 72~GeV^2~.
\label{kinreg}
\end{equation}

As in our previous analyses \cite{LSiSt98, schemedep}
the total (statistical and systematic) errors
are taken into account. The results presented in Table 1
correspond to an SU(3) symmetric sea. Note that in this case 
$a_{\bar{s}}=a_S$ and the
first moment of the strange sea quarks, 
$\eta_{\bar{s}}\equiv \Delta\bar{s}(1,Q^2_0)=\eta_S/6~$. 

It is seen from the Table 1 that the values of $~\chi^2/DOF~$ 
coincide almost exactly in the different factorization schemes, 
which is a good indication of the stability of the analysis.
The NLO QCD predictions are in a very good
agreement with the presently available data on $A^N_1$ and $g^N_1$,
as is illustrated in the JET scheme fit in Figs. 1a-c.
\vskip 0.6 cm
\begin{center}
\begin{tabular}{cl}
&{\bf Table 1.} Results of the NLO QCD fits in the JET, AB and 
$\overline{\rm MS}$ schemes to\\ 
&the world $~A_1^N~$ data ($Q^2_0=1~GeV^2$). The errors shown 
are total (statistical\\ 
&and systematic). 
\end{tabular}
\vskip 0.6 cm
\begin{tabular}{|c|c|c|c|c|c|c|} \hline
 ~~Scheme~~&~~~~~~~~~~~JET~~~~~~~~~~~&~~~~~~~~~~~AB~~~~~~~~~~~
 &~~~~~~~~~~~$\overline{MS}$~~~~~~~~~~~\\ \hline
 $DOF$    &  161~-~6    &     161~-~6   &   161~-~6\\
 $\chi^2$        &  128.3    &  128.4  &   129.4  \\
 $\chi^2/DOF$    &  0.828    &   0.828  &   0.835   \\  \hline
 $a_u$           &~~0.276~~$\pm$~~0.030~~ &~~0.278~~$\pm$~~0.031~~
 &~~0.257~~$\pm$~~0.024 \\
 $a_d$           &~~0.077~~$\pm$~~0.135~~&~~0.065~~$\pm$~~0.136~~
 &~~0.179~~$\pm$~~0.107\\
 $a_{\bar{s}}$   &~~1.524~~$\pm$~~0.391~~&~~1.668~~$\pm$~~0.403~~
 &~~0.761~~$\pm$~~0.188\\
  $\eta_{\bar{s}}$&-~0.032~~$\pm$~~0.005~~&-~0.029~~$\pm$~~0.006~~ 
 &-~0.051~~$\pm$~~0.006\\
 $a_g$           &~~0.175~~$\pm$~~0.452~~&~~0.144~~$\pm$~~0.417~~
 &~~~~3.6~~$\pm$~~3.8\\
 $\eta_g$        &~~~0.57~~$\pm$~~0.14~~&~~0.58~~$\pm$~~0.12~~ 
 &~~~0.07~~$\pm$~~0.10~~\\ \hline
 $\Delta \Sigma(1)$&~~0.389~~$\pm$~~0.037~~&~~0.407~~$\pm$~~0.044~~
&~~0.275~~$\pm$~~0.044\\ 
 $a_0(1~GeV^2)$&~~0.26~~$\pm$~~0.05&~~0.27~~$\pm$~~0.05~~
 &~~0.28~~$\pm$~~0.04\\  \hline
\end{tabular}
\end{center}
\vskip 0.6 cm

Let us now comment briefly upon the state of our knowledge about
the individual polarized parton densities. The extracted quark (valence 
and sea) and gluon densities at $Q^2_0=1~GeV^2$ are shown in Fig. 2 and
Fig. 3, respectively.\\

$\bf {\Delta u_v(x, Q^2)}$ and $\bf {\Delta d_v(x,Q^2)}$

~~i) The valence quark densities are practically unchanged if the 
new (final SMC, HERMES proton and SLAC/E155 deuteron) data are 
incorporated in the analysis. (Note that $\Delta u_v$ is very well 
determined once an assumption about the flavour decomposition of the 
sea is made.)

~ii)$~\Delta u_v$ and $\Delta d_v$ in the JET, AB and 
$\overline{\rm MS}$ schemes coincide within the errors, in excellent
agreement with what follows from the theory (they should be the
same in the factorization schemes under consideration).

iii) Although the first moment of $\Delta d_v(x,Q^2_0),~\eta_d$,  
has been kept fixed in the analysis (see Eq. (\ref{etaudSU3})),
the parameter $a_d$, and therefore $\Delta d_v$ is not well
determined from the present data (see Fig. 2).\\

$\bf {\Delta\bar{s}(x,Q^2)}$ and $\bf {\Delta \Sigma(x,Q^2)}$

Note that these quantities are scheme dependent.

~~i)$~\Delta\bar{s}$ and $\Delta \Sigma$ are well determined from 
the data now (see Fig. 2).

~ii) The first moments $\eta_{\bar{s}}$ in the JET and AB schemes are 
in a very good agreement. The same is valid for the first moments  
$\Delta \Sigma(1)$ (see Table 1). (We recall that according
to the definition of the JET and AB schemes $\Delta \Sigma(1)$
as well as $\eta_{\bar{s}}$ should be the same in both schemes.) 
The corresponding densities
$\Delta\bar{s}(x,Q^2)$ and $\Delta \Sigma(x,Q^2)$ in both schemes
are slightly
different because their higher moments are not equal.

iii) The first moment of $\Delta\bar{s}$ in the $\overline{\rm MS}$
scheme,$~(\eta_{\bar{s}})_{\overline{\rm MS}}$, is almost twice as big as
it is in the JET(AB) schemes, $(\eta_{\bar{s}})_{\rm JET(AB)}$:
\begin{equation}
(\eta_{\bar{s}})_{\overline{\rm MS}}=-0.051\pm 0.006,~~~~~
(\eta_{\bar{s}})_{JET}=-0.032\pm 0.005.
\label{etas}
\end{equation}

The polarized strange quark density is significantly different from zero
independently of the factorization schemes used in the analysis.
The result (\ref{etas}) demonstrates the extent to which the strange
quark densities in different schemes can differ in the polarized case.\\ 

{\bf Gluons $\bf \Delta G(x,Q^2)$}

~~i) The new set of data allows a better determination of the
polarized gluon density in the JET and AB schemes.

In previous studies $\chi^2$ was quite insensitive to the value
of $a_g$ in all factorization schemes under consideration, and
the results of the fits corresponding to fixing $a_g=0.6$ were
presented in \cite{LSiSt98, schemedep}. The results when $a_g$ 
is a free parameter are summarized in Table 2.  
\vskip 0.6 cm
\begin{center}
\begin{tabular}{cl}
&{\bf Table 2.} The values of the parameters associated with the input\\
&polarized gluon density in the JET, AB and $\overline{\rm MS}$ 
schemes for the\\
&new and old sets of data.
\end{tabular}
\vskip 0.6 cm
\begin{tabular}{|c|c|c|c|c|c|c|} \hline
Param.&~Data set~&~~~~~~~JET~~~~~~~&~~~~~~~AB~~~~~~~
&~~~~~~~$\overline{\rm MS}$~~~~~~~\\ \hline
$a_g$& old &~~1.8~$\pm$~2.1~~ &~~1.7~$\pm$~1.8~~
&~~1.6~$\pm$~1.9 \\
$\eta_g$& old &~~0.24~~$\pm$~~0.35~~&~~0.28~~$\pm$~~0.40~~
&~~0.30~~$\pm$~~0.50\\ \hline
$a_g$& new &~~0.18~~$\pm$~~0.45~~&~~0.14~~$\pm$~~0.42~~
&~~3.6~$\pm$~3.8\\
$\eta_g$& new &~~0.57~~$\pm$~~0.14~~&~~0.58~~$\pm$~~0.12~~ 
&~~0.07~~$\pm$~~0.10\\  \hline
\end{tabular}
\end{center}
\vskip 0.6 cm

It is clear from the Table 2 that now both the value of $a_g$
and $\eta_g$, i.e., the shape and the normalization of the gluon 
density are much better constrained in the JET and AB schemes. 
The progress in extracting the polarized gluon density from the
new data set is illustrated in Fig. 3 (JET scheme).

~ii) In the $\overline{\rm MS}$ scheme the gluons are still
poorly constrained. $\chi^2$ continues to be insensitive to a large
range of possible values for $a_g$ (see Table 2). 
The obtained value of $\eta_g$ in this case is consistent with zero.
However, if we fit the data with a fixed value of $a_g=0.2$
(in accord with the values obtained in the JET, AB schemes),
we find $\eta_g=0.62+/-0.42$, in agreement with the values
of this quantity in the JET and AB schemes, as expected theoretically.\\

{\bf Axial charge $\bf {a_0(Q^2)}$} 

~i) The good agreement between the values of $a_0(Q^2)$ 
determined in the different schemes is
demonstrated in Table 1, which illustrates how our analysis
respects the scheme-independence of physical quantities. 

ii)The central values of $a_0$ at $Q^2 = 1~GeV^2$ are slightly 
smaller than the corresponding values obtained from our previous 
analysis.\\

We have also investigated the sensitivity of $\chi^2$ to the
alternative possibilities of explanation of the spin of the
nucleon via the spins of its constituents, namely:

~~i) $~\eta_{\bar{s}}\neq 0,~\eta_g= 0~$, so that 
$a_0 \approx \Delta \Sigma~<<~a_8 = 0.58$

~ii) $~\eta_{\bar{s}}=0,~\eta_g \neq 0~$, so that
$a_0~<<~\Delta \Sigma \approx a_8 = 0.58$

iii) $~\eta_{\bar{s}} \neq 0,~\eta_g \neq 0~$\\

The numerical results (JET scheme) are given in Table 3.
\vskip 0.6 cm
\begin{center}
\begin{tabular}{cl}
&{\bf Table 3.} Sensitivity of $\chi^2$ to the absence of polarized
strange\\ 
&quarks $\Delta\bar{s}$ or gluons $\Delta G$ at $Q^2 = 1~GeV^2$.
\end{tabular}
\vskip 0.6 cm
\begin{tabular}{|c|c|c|c|c|c|c|} \hline
 ~~Assumption~~&~~$\chi^2$~~&$-~\eta_{\bar{s}}$ &
 $\eta_g$ & $a_0$ \\ \hline
 $~\eta_{\bar{s}}\neq 0,~\eta_g= 0~$ & 133.1 &0.065~$\pm$~0.014 &~0~&
 0.19~$\pm$~0.09\\
 $~\eta_{\bar{s}}=0,~\eta_g \neq 0~$ & 134.7 &~0~ & 
 1.33~$\pm$~0.11 & 0.27~$\pm$~0.04\\
 $~\eta_{\bar{s}} \neq 0,~\eta_g \neq 0~$ & 128.3 & 0.032~$\pm$~0.005 & 
 0.57~$\pm$~0.14 & 0.26~$\pm$~0.05\\ \hline
\end{tabular}
\end{center} 
\vskip 0.6 cm

It is seen from the Table that the present inclusive DIS data 
prefer small strange
quark and small gluon polarized densities at $Q^2 = 1~GeV^2$:
\begin{equation}
\eta_{\bar{s}}= -0.032\pm 0.005,~~~\eta_{g}=0.57\pm 0.14.
\label{etasg}
\end{equation}
\newpage
{\bf First moments of the spin structure functions}

Finally, we present our results for the first moments of the
nucleon spin structure functions $g_1^N$ at $Q^2 = 5~GeV^2$ in
the measured $x$ range from 0.003 to 0.8 (AB scheme):
\begin{eqnarray}
\nonumber
&&~~0.133~~~{\rm for~Proton}\\
\int _{0.003}^{0.8}dxg_1^N(x,Q^2=5~GeV^2)_{fit}~~=&&~~0.039~~~{\rm for~
Deuteron}\\ 
\nonumber
&&-0.048~~~{\rm for~Neutron}
\label{Gamma1}
\end{eqnarray}
which are in excellent agreement with their experimental values
\cite{SMC/QCD}
obtained from the world set of data (HERMES proton and E155
deuteron data not included):
\begin{eqnarray}
\nonumber
&&~~0.130~\pm~0.003(stat)~\pm~0.005(syst)~\pm~0.004(evol)
~~~~{\rm for~Proton}\\
&&~~0.036~\pm~0.004(stat)~\pm~0.003(syst)~\pm~0.002(evol)
~~~~{\rm for~Deuteron}\\
\nonumber
&&-0.054~\pm~0.007(stat)~\pm~0.005(syst)~\pm~0.004(evol)
~~~~{\rm for~Neutron}
\label{Gamma1exp}
\end{eqnarray}

{\bf The nucleon spin}

In the JET and AB schemes it is meaningful to interpret 
$\Delta \Sigma(1)$, the first moment of the singlet density
$\Delta \Sigma(x,Q^2)$, as the contribution of the quark spins 
to the nucleon's spin. Our value of 
$~\Delta \Sigma = 0.40\pm 0.04~$ is not far from the value 
0.6 expected in certain quark models \cite{JaffeManohar}.
For the spin contribution from both quarks and gluons we have
(JET scheme):
\begin{equation}
{1\over 2}\Delta \Sigma(1) + \eta_g = 0.76\pm 0.14
\end{equation}

This value is consistent with 1/2 in two standard deviations. 
The more accurate determination of $\Delta \Sigma$ and $\eta_g$ 
will answer the basic question how the spin of nucleon is 
divided up among the spin of quarks and gluons and their 
orbital angular momenta.\\

In conclusion, we have re-analyzed the world data on inclusive 
polarized deep inelastic lepton-nucleon scattering in NLO QCD 
adding to the old set of data the final SMC results, the new
HERMES proton and SLAC/E155 deuteron data. As in the previous
analysis it was
demonstrated that the polarized DIS data are in an excellent
agreement with the pQCD predictions for $~A_1^N(x,Q^2)~$ and
$~g_1^N(x,Q^2)~$ in all 
the factorization schemes considered. The polarized parton
densities have been extracted from the data. We have found that
the main effect of the newly incorporated data is a better
determination of the polarized gluon density, but in comparison
with the other densities the uncertainty is still large.
It follows from our analysis that the present inclusive DIS data
prefer small strange quark and small gluon polarized  densities, 
but they are significantly different from zero.

To test more precisely the spin properties of QCD and to determine
better the polarized densities, accurate data from both the
inclusive (neutral and charged current) polarized DIS and the
semi-inclusive processes in a larger kinematical region are needed.
Finally, a direct measurement of the gluon polarization is necessary.
We hope the COMPASS experiment at CERN, the future experiments at the
hadron collider RHIC and the possibility of having a polarized proton 
beam at HERA will help to further a more profound study of the
internal structure of the nucleon.\\

\vskip 3mm
This research was partly supported by a UK Royal Society Collaborative
Grant, by the Russian Foundation for Basic Research, Grant No 
99-01-00091 and by the Bulgarian 
science Foundation under Contract \mbox{Ph 510.}\\

\newpage
\noindent
{\bf Figure Captions}
\vskip 3mm
\noindent 
{\bf Fig. 1.} Comparison of our NLO results in the JET scheme
for $~A_1^N(x,Q^2)~$ (a,b) and $~xg_1^N(x,Q^2)~$ (c) with the 
experimental data at the measured $x$ and $Q^2$ values. 
Errors bars represent the total error.\\

\noindent
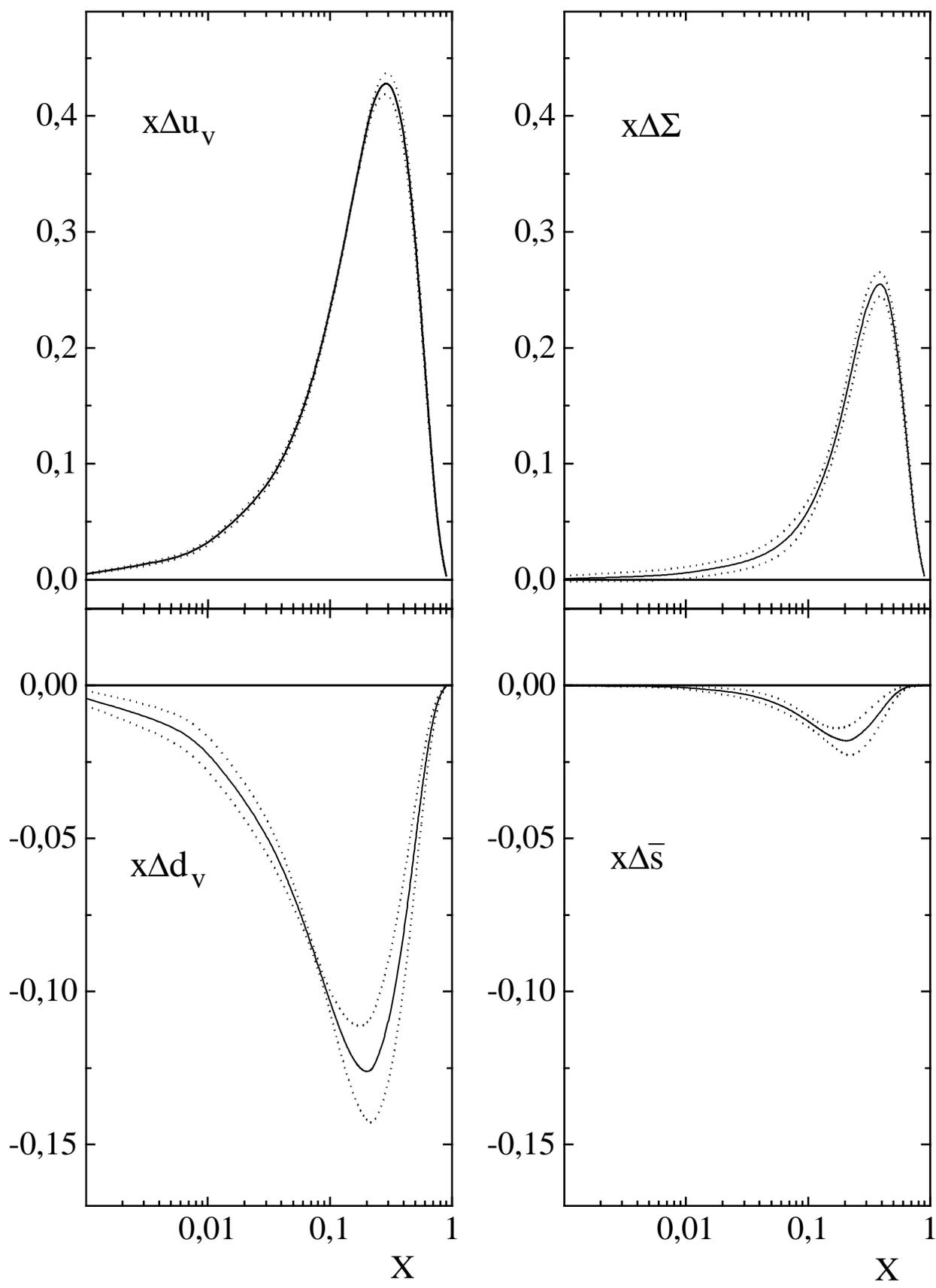
{\bf Fig. 2.} Next-to-leading order polarized valence, strange
and singlet quark distributions at $~Q^2=1~GeV^2~$ with their
error bands (JET scheme). The systematic errors are added 
quadratically.\\

\noindent
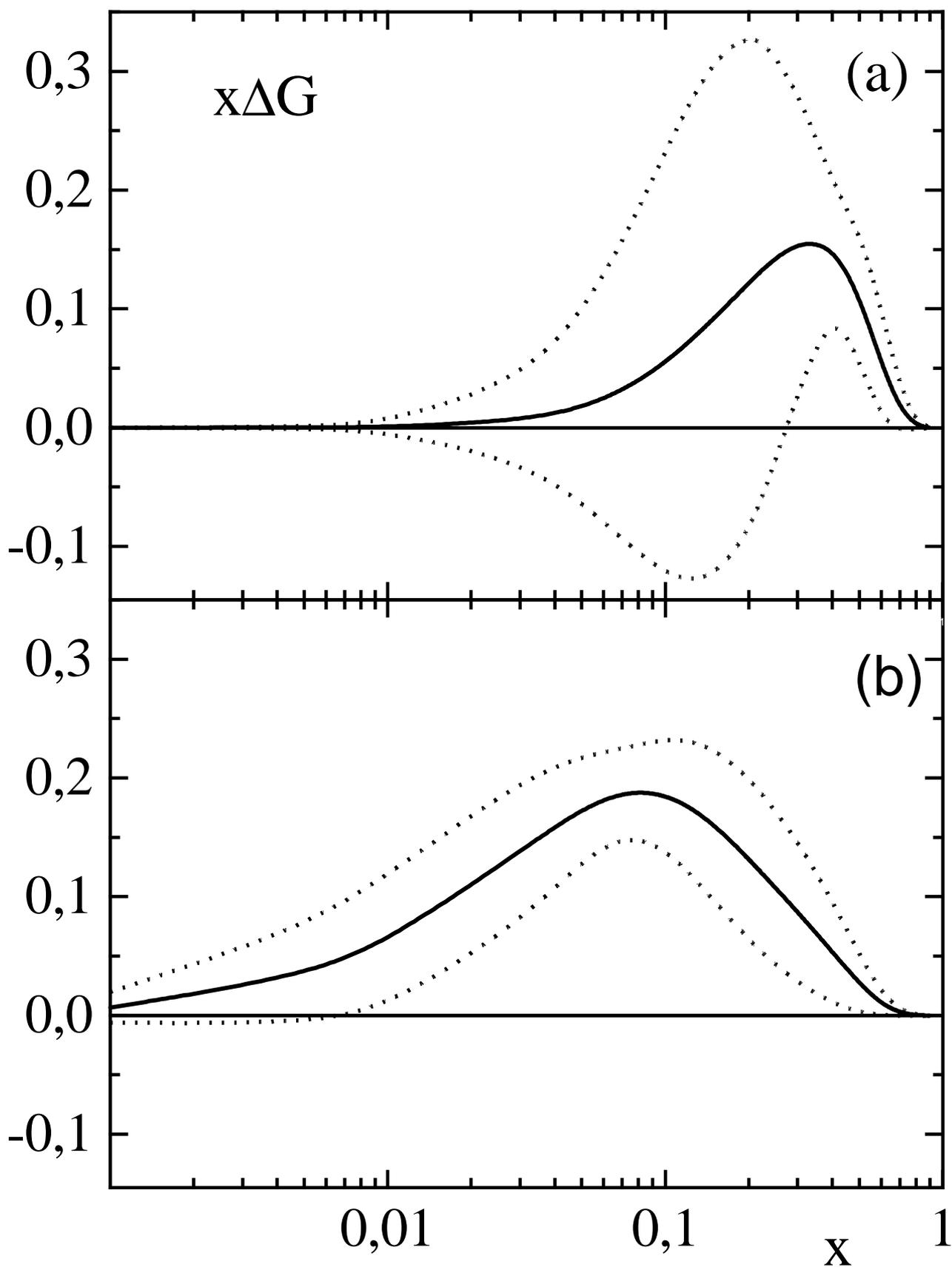
{\bf Fig. 3.} Next-to-leading order polarized gluon
density at $~Q^2=1~GeV^2~$ determined
from the old (a) and the new (b) world sets of data
(JET scheme). The error bands account for the statistical 
and systematic uncertainties.

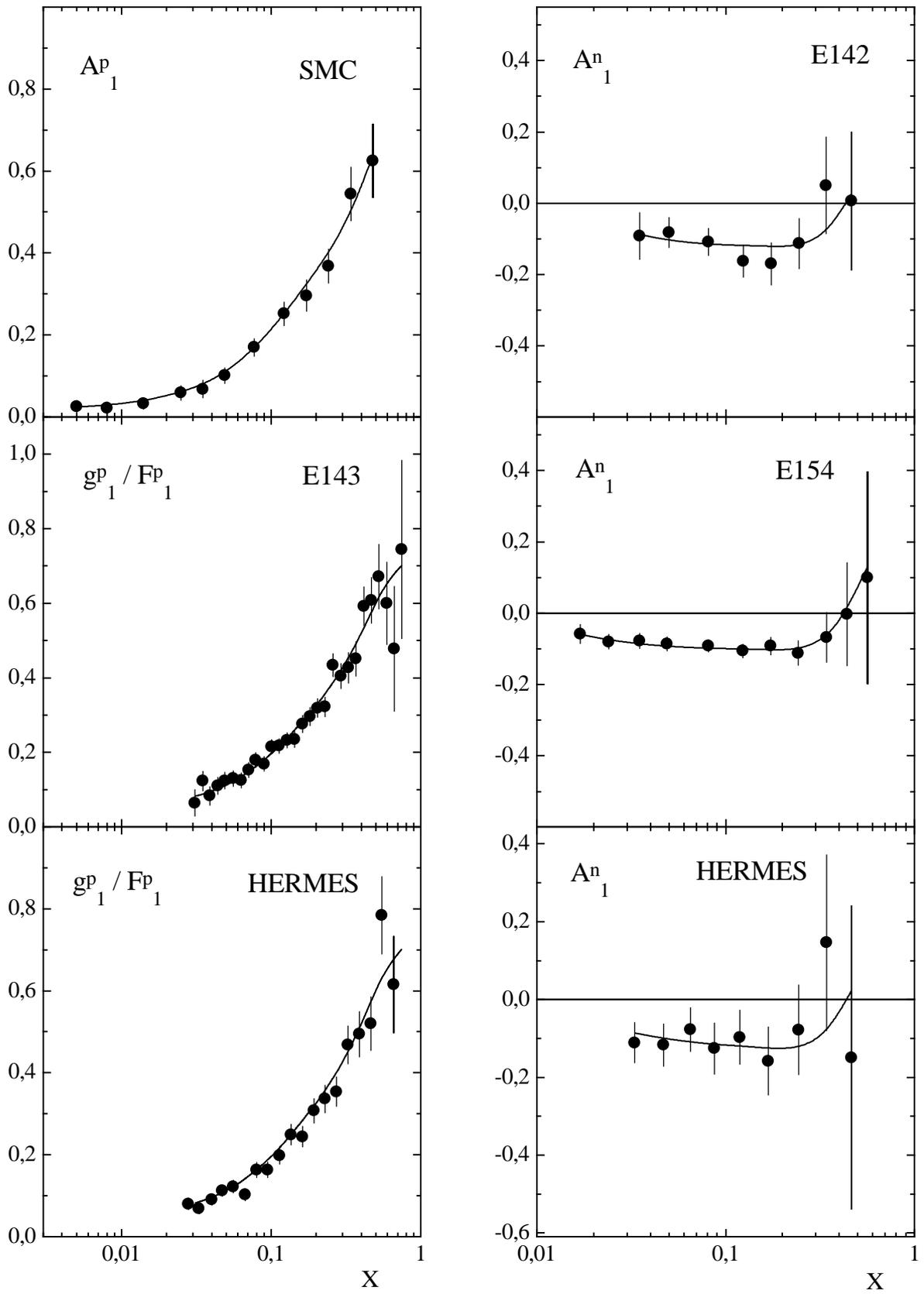
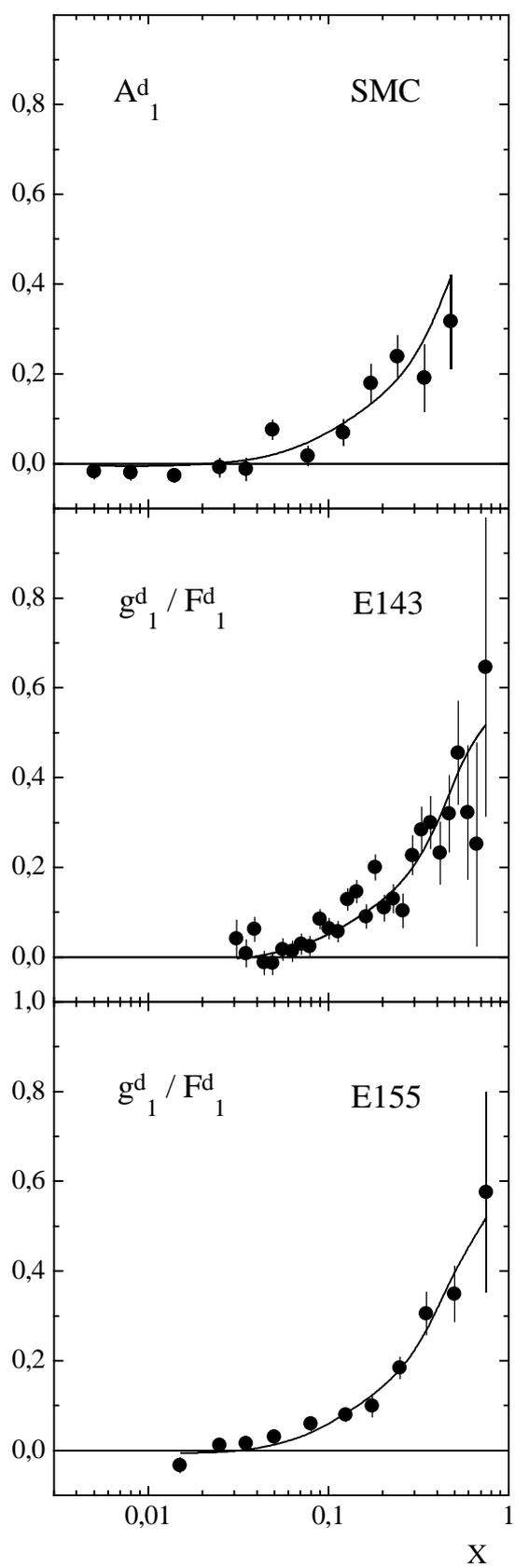
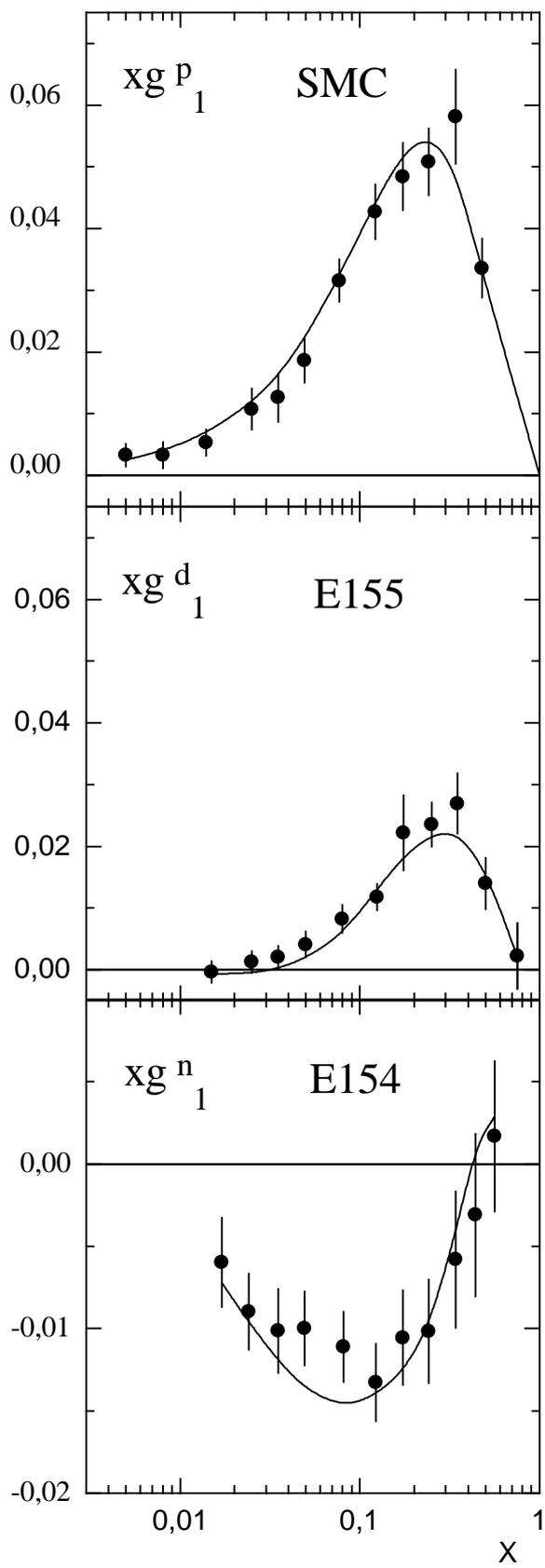
\end{document}